\documentclass[mathleft]{an}
\usepackage{graphicx}
\usepackage{times}
\begin{document}

% The following seven commands are intended for editorial usage and should be ignored by
% the author(s).
\Pagespan{789}{}% Document's page range. 
% If second parameter is left empty, the last page is computed automatically.
\Yearpublication{2006}%
\Yearsubmission{2005}%
\Month{11}%   
\Volume{999}%  
\Issue{88}% 
% \DOI{This.is/not.aDOI}% 

\title{The shapes of light curves of Mira-type variables}

\author{T. Lebzelter\inst{1}\fnmsep\thanks{Corresponding author:
  \email{thomas.lebzelter@univie.ac.at}\newline}
%Example 
%for footnote, note the usage of the \texttt{fnmsep}
%command as separator between institute number and footnote mark} 
}
\titlerunning{Shapes of mira-type light curves}
\authorrunning{T. Lebzelter}
\institute{
University of Vienna, Department of Astronomy, 
T\"urkenschanzstrasse 17, 1180 Vienna, Austria
}

\received{20 July 2010}
\accepted{?}
\publonline{later}


\abstract{%
Using a sample of 454 mira light curves from the ASAS survey we study the shape of the light variations
in this kind of variable stars. Opposite to earlier studies, we choose a general approach to identify any deviation
from a sinusoidal light change. We find that about 30\% of the studied light curves show a significant
deviation from the sinusoidal reference shape. Among these stars two characteristic light curve shapes
of comparable frequency could be identified. Some hint for a connection between atmospheric chemistry
and light curve shape was found, but beside that no or only very weak relations between light curve shape
and other stellar parameters seem to exist.}

\maketitle

\section[]{Introduction}
Mira (o Cet) is the longest known periodic variable. The variability class represented by this object, dubbed
miras, is characterized by large amplitude (several magnitudes in $V$) and long period ($>$ 100\,d) 
variations. From the point of
stellar evolution, miras are highly evolved stars of low or intermediate mass that are found on the 
Asymptotic Giant Branch (AGB) in the Hertzsprung-Russell diagram (HRD). Their high luminosity allows to observe them throughout the Milky
Way and the Local Group (e.g. Groenewegen 2004), and even beyond (e.g. Rejkuba 2007). Most of them are radial,
fundamental mode pulsators (e.g. Wood \& Sebo 1996).

Despite the long history of observations of miras and the large number of light curves obtained -- with
a key contribution coming from amateur astronomers (AAVSO) -- 
systematic studies on the light curve shape of these stars are 
rare. This lack is somewhat surprising as several papers pointed out that there seems to be a relation between the 
shape of the visual light curve and various other parameters, mostly related to the circumstellar material around 
these stars: Bowers (1975) and Bowers \& Kerr (1977) were probably the first to notice that the occurrence of 
microwave OH emission in miras seems to be linked to the steepness of the rising branch of the visual light curve. 
Vardya et al. (1986) found that the appearance of silicate emission features at 9.7 and 20 $\mu$m is related to the 
asymmetry of the light curve. This was later confirmed by Onaka et al. (1989). It was again Vardya (1987), who showed 
that the probability of detecting the H$_{2}$O vapor line at 1.35 cm depends on the light curve shape. Le Bertre 
(1992) and Winters et al. (1994) explored a possible relation between light curve shape and dust formation.
Mennessier et al. (1997) discriminated M- and C-stars by using period, amplitude, light asymmetry and IRAS
colours.
Existing and forthcoming large datasets of light curves of long period variables from groundbased photometric
surveys thus motivate a closer look on the possibility to derive basic stellar quantities like evolutionary status or 
mass loss properties from light curve data. For this, an easily applicable and quantifiable method to describe the  
light curve shape is needed.

Special features in the light curve shape of these variables have been noted by several authors. Humps on the rising 
branch of the light curve as well as a clear asymmetry between the rising and descending branch have been noted e.g. 
by Lockwood \& Wing (1971), also summarizing results from earlier studies. But it is also clear that only a fraction 
of all miras show this behaviour. The humps may be analogs of the humps seen on bump Cepheid light curves and likely 
result from a 2-to-1 resonance between the fundamental and the first overtone mode (Barthes et al. 1998, Wood et al. 
1999, Lebzelter et al. 2005). A few miras have been reported to show even double maxima in their light curve (e.g. 
Keenan et al. 1974). Comparison with infrared colour-period relations suggest that the true period is actually half 
the period between deep minima (i.e. only one maximum per period, cf. Feast et al. 1982) in these cases. 

Early work on the classification of mira lightcurves
has been summarized by Payne-Gaposchkin \& Gaposchkin (1938), pointing out that for correlations
of light curve shape with other parameters a numerical value is in advantage over a formal classification
system. 
The most extensive study of the shape of mira light curves up to now is the work of Vardya (1988) who analysed light 
curve data of 368 miras from the literature. Their paper also outlines the two major approaches used to describe the 
shape of the light curve, namely on the one hand the qualitative classification scheme given by Ludendorff (1928) and 
on the other hand the visual light asymmetry factor {\it f}, which is defined as the rise time over the period (in 
days). As a result of this study, Vardya found that 80\,\% of the miras lie in the range 0.4$\le f \le$0.5, i.e. 
only a small fraction of the miras show strong deviations from a symmetric visual light curve. This fraction seems to 
be dependent on the atmospheric chemistry (M-, S- or C-stars). Vardya also compared the $f$ values with the classes 
defined by Ludendorff, but no clear correlation could be found. This is also hampered by the fact that the classes of 
Ludendorff are not very clearly defined and the classification seems to be somewhat subjective. However, there is 
also some uncertainty in the $f$-parameter. As pointed out by Onaka et al. (1989)
this uncertainty may reach a value of 0.1. Therefore, the description of the light curve shape and accordingly the 
frequency of asymmetries has to be seen as not satisfyingly solved (as was also
noted by Vardya 1988). 

The study of Vardya is based on the archive of observations from amateur astronomers
(e.g. Campbell 1955). In contrast, nowadays a large number of photometric light curves in the visual range are 
available from various surveys like MACHO (Wood et al. 1999) or 
OGLE (Soszynski et al. 2009, survey mainly done in the $I$-band). 
These data have led to major advances in our 
understanding of long period variables, especially in terms of the period-luminosity relations (e.g. Wood et al. 
1999, Ita et al. 2004, Soszynski et al. 2007) or the study of 
long secondary periods (e.g. Wood 2000, Soszynski et al. 2007, Wood \& Nicholls 2009). 
However, a systematic study of these data sets in 
terms of light curve asymmetry has not been done
yet.

In this paper the deviation of the shape of the light curves of miras from a sinusoidal variation is 
investigated for a selected sample of sources in the All-Sky Automated Survey (ASAS 3) variable stars
catalogue (Pojmanski 2002). 
Opposite to other surveys like MACHO or OGLE, ASAS is focusing on the variability of comparably bright field stars. 
Its detection limit is close to a $V$ magnitude of 14. Photometric accuracy is given to be about 0.05 mag, but may be 
somewhat worse in some cases. Despite these shortcomings the ASAS catalogue was selected for our study as many of 
these bright targets have additional information available, e.g. from 2MASS, IRAS or light curves from other sources.

\section[]{Sample selection}
Starting point for our selection was the ASAS catalogue of variable stars (ACVS). All ASAS data 
including this catalogue are 
accessible via the ASAS webpage\footnote{http://www.astrouw.edu.pl/asas/}.
The variability type found in that catalogue has been assigned by means of an
automated classification scheme using both light curve parameters (period, amplitude and
Fourier coefficients) and near infrared colours (taken from 2MASS). See Pojmanski (2000, 2002) 
and Pojmanski \& Maciejewski 
(2004) for details. In total 2895 objects in the catalogue are classified as miras. We note here that about 850
of the stars classified as miras in the ACVS have a $V$ amplitude of less than 2.5 magnitudes. 
According to the classical scheme (Kholopov 1985-88, GCVS) such stars would be rather classified as SRa. For the 
beginning we did not exclude these stars from our sample, but we will later come back to this point.

A quick visual check of all selected light curves was done. During this step, the period determination of several 
stars was slightly improved and a small number of stars were rejected due to an unclear variability pattern or an 
obviously low photometric accuracy of the data. In the next step,
the time series were transformed into phased light curves ([0,1])
using the revised periods from the previous step. Each of these light curves was then split into steps of 0.05 in 
phase, and the data within each slot were averaged. At this step we kept only those stars that had no gaps in their 
phased light curves, i.e. none of the slots was empty. Naturally, we lost most stars with periods close to one year 
in this step, because such objects were always unobservable at the same phase. 
As a result we had averaged light curves with a step size of 0.05 in phase.

The resulting sample included a remarkable fraction of light curves with extremely broad minima. According
to Pojmanski (priv.~comm.) such a behaviour may be mimicked by blending with nearby stars. The limited 
spatial resolution of ASAS does not allow to separate stars closer than 15 arcsec. If the minimum of the mira is at a 
lower brightness than the blending star, the resulting light curve would show such a broad minimum. To check for this 
possible disturbance we searched the USNO A2 catalogue for
stars  within 20 arcsec of the sample miras that show an $R$-magnitude brighter than 15. Removal of
these sample stars reduced the number of light curves with such broad minima to very few cases and the total sample 
to approximately 500 objects, from which 454 stars were then chosen as the final sample (removing some cases of 
uncertain or possibly variable period or a large scatter of the averaged data points of the mean light curve).

\begin{figure}
\includegraphics[width=80mm]{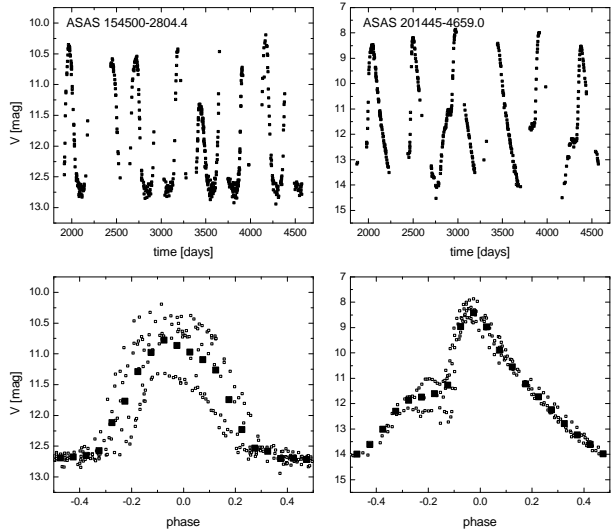}
\caption{Illustration of the approach used to derive mean light curves. The top row shows the light change
of two example stars against time. The bottom row shows the observed data against phase (small open symbols)
and the resulting average light curve (solid symbols). The example on the right has been used in the further
analysis, the example on the left was rejected due to large scatter and blending with a nearby star
(flat minimum).}
\label{method}
\end{figure}

To illustrate the method we show in Fig.\,\ref{method} two examples of mira lightcurves from the ASAS archive.
The top row shows the photometry against time directly taken from the archive. It is obvious that the left
example (ASAS 154500-2804.4) shows quite strong variations of the amplitude, while in the example on the right
a bump can be seen on the rising branch of the light change (ASAS 201445-4659.0). The row below now gives the
phased light curve (small open symbols) and on top of it the averaged lightcurve for each of the two objects
(filled large symbols). The left example has been rejected from our final sample for two reasons: first, the average of the measured standard deviation of all data points in each bin exceeds a limit of 0.6 magnitudes\footnote{It is
irrelevant whether the scatter is produced by amplitude or period changes or by short time 
irregularities in the light change.}. We did make
some experiments with slightly changing this limit, but the effect on the results was minor. We note that we
rejected also stars with a very large scatter (exceeding 1 magnitude) in a single bin. Second, the light curve
shows a very flat and broad minimum indicating blending with a nearby star as discussed above. The light curve 
shown on the right side of Fig.\,\ref{method} shows much less scatter. While the bump on the rising branch
is clearly supported by observations from several light cycles, cycle-to-cycle variations are present. The mean
light curve is describing the light change outside the bump very well, and is also representative for an 
average shape of the bump itself as one would fit it by eye.

Taking into account the fact that mira pulsation is not strictly periodic the derived average light curve has
to be seen as a typical light change during the time of the observations. For stars with a very short observing
window at the site of the ASAS telescope (Las Campanas Observatory, Chile) and thus a bad phase coverage,
asymmetries could be smeared out or even falsely introduced by cycle-to-cycle variations. We tried to carefully check
those light curves that may be affected by this problem and removed unclear cases.

\begin{figure}
\includegraphics[width=80mm]{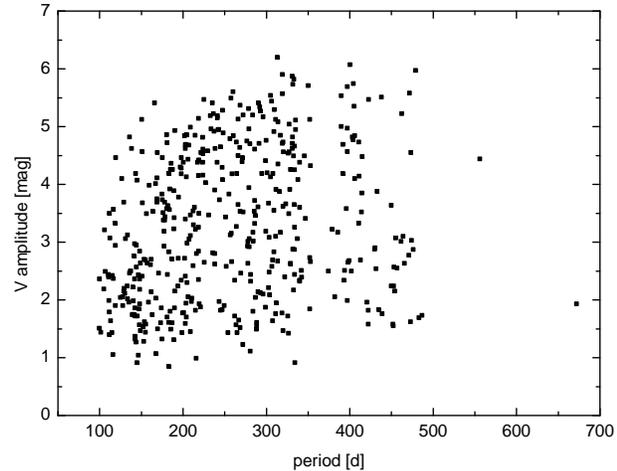}
\caption{Distribution of the sample stars in the visual light amplitude vs. period plane.}
\label{PvsVamp}
\end{figure}

Figure \ref{PvsVamp} shows the distribution of the sample stars in the period and amplitude plane. The typical period 
range of miras between 200 and 500 days is well covered. As mentioned above, we have a gap for periods around 1 year 
as the result of our selection process. It is also clear, that our sample includes several stars with amplitudes 
between 1 and 2 mag, i.e. these objects have a too small amplitude for miras in the classical scheme. Short period or 
small amplitude semiregular variables are absent from our sample, furthermore we have only single period objects in 
our study.

\section[]{Light curve analysis}
For the analysis we decided to choose a different approach for describing the shape of the light curve than previous 
investigations: the light curve was compared to a sinusoidal reference curve, and the sum of the squared differences 
was taken as a measure for the deviation. For this purpose we first transformed the
averaged light curves of our sample to a brightness range between 0 (brightest point) and 1 (weakest point). This 
transformed light curve (in the following we will call this product {\it normalized averaged light curve})
was then compared to a sinusoidal reference curve. The comparison was done by calculating the value 

$\chi^2=\sum_{i=1}^{20}{(X_O(\phi_i) - X_S(\phi_i))^2}$,

 where $X_O$ and $X_S$ are the 20 individual data 
points of the normalized averaged light curve and the sinusoidal reference curve, respectively, and $i$ is
an index running from minimum phase to minimum phase. 
The light curve of the target was
shifted relative to the sinusoidal reference curve until a minimum value for the squared difference was reached,
which was used for the further analysis. 
This was done to compensate for possible phase shifts.

\begin{figure}
\includegraphics[width=80mm]{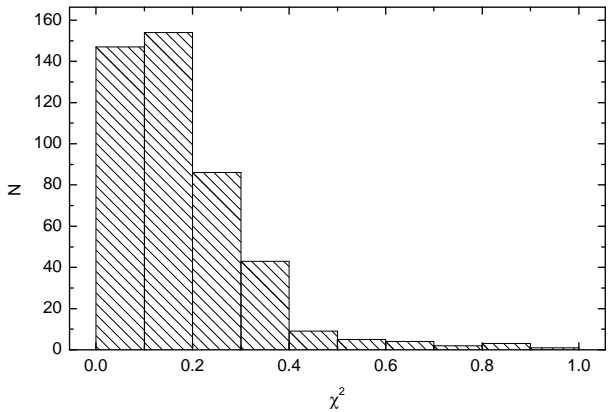}
\caption{Histogram of the differences ($\chi^{2}$) between averaged normalized light curve and
reference curve found for our sample of mira-like variables. Small values indicate a variation
close to sinusoidal, while large values mark stars strongly deviating from a sinusoidal light change.}
\label{chi2distrib}
\end{figure}

\begin{figure*}
\includegraphics[width=160mm]{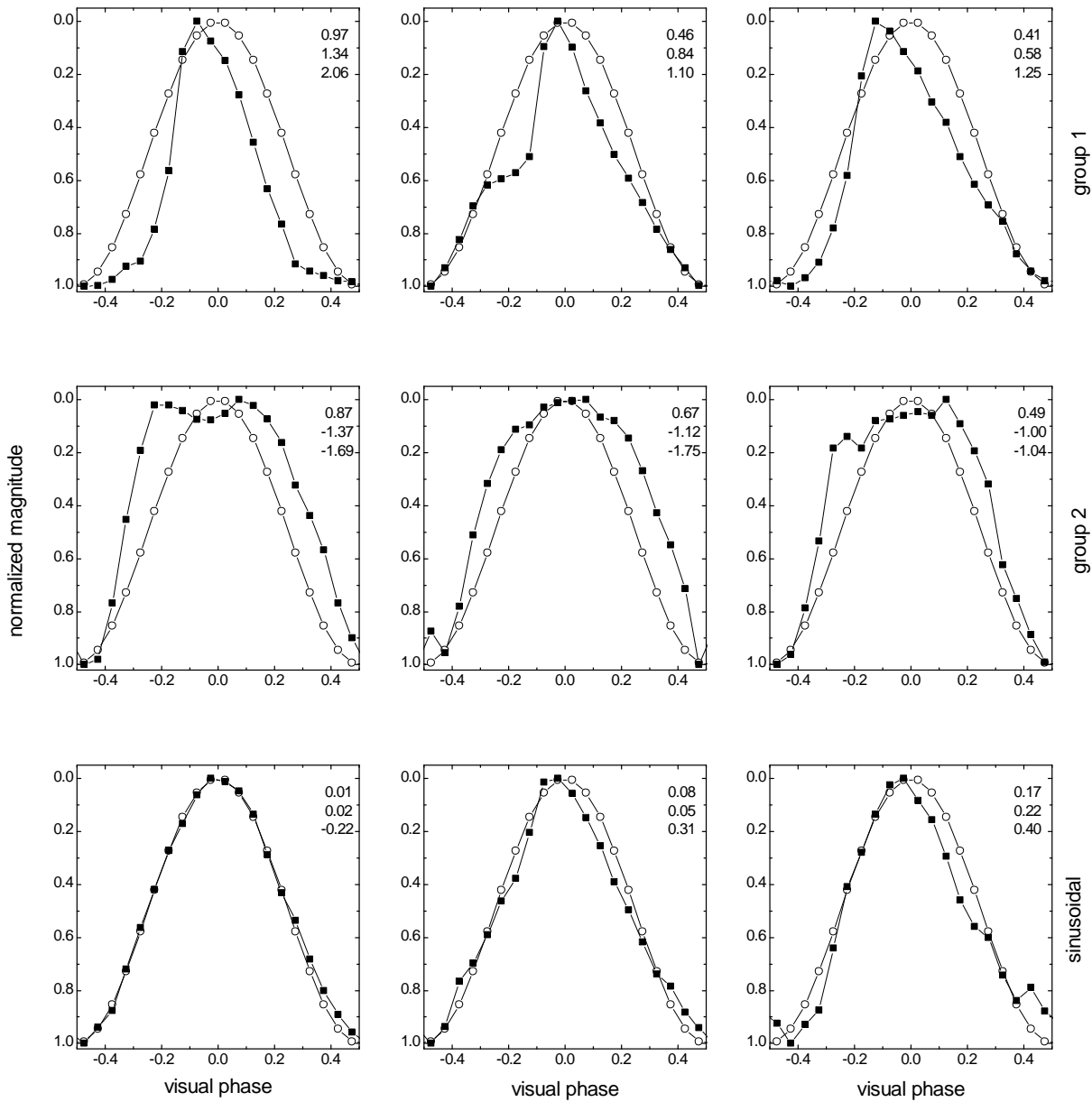}
\caption{Examples for normalized averaged light curves of miras from the ASAS variable star catalogue. A value of
0 on the y-axis
corresponds to light maximum, while a value of 1 marks the minimum. The observed data are indicated by filled
symbols, while the sinusoidal comparison curve is marked by open symbols. 
For illustrative purpose the first and last data point of each light curve has been repeated and the phase
of the light maximum ($\Phi$=0) has been shifted to the centre of each panel. The numbers in
the upper right corner of each panel gives the corresponding value of $\chi^{2}$, $\chi_{1}$ and
$\chi_{2}$, respectively.
Top row (group 1): 
ASAS 171611-6148.0 (DD Ara), ASAS 201445-4659.0 (R Tel), ASAS 202916-4025.1 (U Mic). Second row (group 2):
ASAS 162522-5827.8 (EQ Nor), ASAS 173706+1813.1 (FR Her), ASAS 205300+2322.3 (RX Vul). Bottom row
(sinusoidal): ASAS 051917-3342.5 (T Col), ASAS 233227-4559.3 (V Phe), ASAS 134821-3651.7 (RT Cen).
}
\label{Examples}
\end{figure*}

Figure \ref{chi2distrib} shows the distribution of the sums of the squared differences ($\chi^{2}$) 
between normalized averaged light curve and
reference curve found for our sample stars. Small values indicate a light change very close to a sinusoidal
curve. Inspection of individual cases reveals that up to a value of $\chi^{2}$=\,0.2 only minor deviations 
from the reference curve can be seen. The transition towards a clearly non-sinusoidal light change is smooth.
The maximum value found for our sample stars was $\chi^{2}$=\,0.97, which
was measured for the mira DD Ara (ASAS 171611-6148.0, 
P=284\,d, $\Delta V$=3 mag). We note that this extreme case shows also some cycle to cycle variations in the 
light amplitude. On the other hand, the smallest value of $\chi^{2}$ was found for the
star T Col (ASAS 051917-3342.5, P=230\,d, $\Delta V$=4.6 mag). Various examples of light curves
are shown in Fig.\,\ref{Examples}.

For a better characterization we also determined the absolute differences between each point of the
normalized averaged light curve and the reference curve and summed them up for the first and the 
second half of the light cycle separately. 
We define $\chi_{1}$ as $\sum_{i=1}^{10}{(X_O(\phi_i) - X_S(\phi_i))}$ and 
$\chi_{2}$ as $\sum_{i=11}^{20}{(X_O(\phi_i) - X_S(\phi_i))}$, respectively. 
As we set the light curve maximum to 0 and the minimum to 1, a positive difference $(X_O(\phi_i) - X_S(\phi_i)$) 
means that the observed curve is fainter at a given phase than what would be expected from a sinusoidal
variation.

Depending
on the deviation of the light curve from the sinusoidal variation -- resulting either in a more narrow or in a
broader peak -- this criterion may allow to distinguish various kinds of light curve shapes. Indeed our results seem
to prove that. Illustrative examples for various cases are shown in Fig.\,\ref{Examples}. 

We can roughly define
two main groups of light curves significantly deviating from a sinusoidal variation: 
\begin{itemize}
\item group 1: $\chi_{1}$ and $\chi_{2}$ are both positive (upper row of Fig.\,\ref{Examples}). This
is the classical case of an asymmetric mira light curve, and it is characterized by a broad and flat 
minimum followed by
a rather steep rise to a narrow maximum. The way back to the minimum is less steep than the rising branch.
Humps may occur on the lower part of the rising branch, in rare cases also on the declining branch.

\item group 2: $\chi_{1}$ and $\chi_{2}$ are both negative (middle row of Fig.\,\ref{Examples}). Here the minimum 
phase
is rather narrow. The rising branch may again be quite steep, but the curve continues into a broad maximum. Special
cases in this group are light curves with two maxima, which may be related to stars showing a hump on the upper
part of the rising branch and which are also found in this group. 
\end{itemize}

Most stars with $\chi^{2}$ values above 0.2 can unambiguously be attributed to either of these groups. The 
bottom row of Fig.\,\ref{Examples} shows some examples of light curves close to a sinusoidal variation. 

\begin{figure}
\includegraphics[width=80mm]{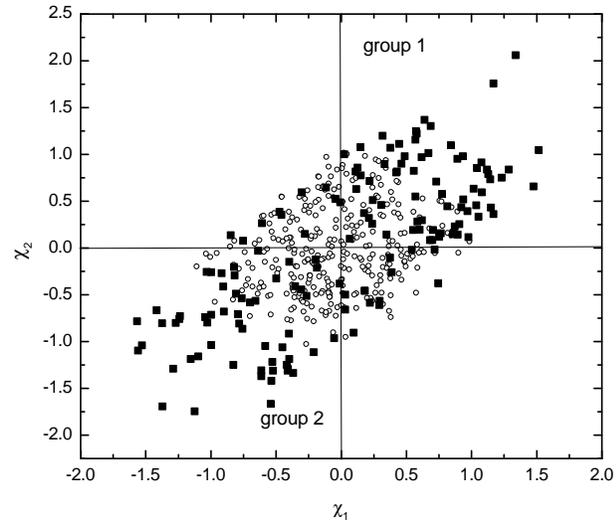}
\caption{Parameters $\chi_{1}$ vs. $\chi_{2}$ for our sample of light curves. Stars with $\chi^{2}$\,$<$\,0.2 
are marked by an open circle, while filled boxes denote objects with $\chi^{2}$\,$>$\,0.2. The two groups
of stars deviating from a sinusoidal variation (see text) are found in the upper right -- group 1 -- and
lower left -- group 2 -- corner, respectively.}
\label{chi1chi2}
\end{figure}

In Fig.\,\ref{chi1chi2} we plot $\chi_{1}$ against $\chi_{2}$. Stars in the upper right corner belong to group 1,
stars in the lower left to group 2. It can be seen that both groups include a similar number of objects. If we
count only the stars with $\chi^{2}$\,$>$\,0.2 (filled 
symbols in Fig.\,\ref{chi1chi2}), we end up with 74 stars (or 16\,\%
of the total sample) in group 1 and 54 objects (12\,\%) in group 2. About 20 stars with 
$\chi^{2}$\,$>$\,0.2 fall outside these two groups. None of them shows extreme $\chi_{1}$ or $\chi_{2}$
values. Checking these stars in detail reveals that they are a mixture of stars that do indeed show
a light curve shape different from group 1 and group 2, and stars that are close to our scatter limit
around the mean light curve (i.e. typically stars with cycle-to-cycle variations). For the first case,
the number of stars is too small for a more detailed study.

\begin{figure}
\includegraphics[width=80mm]{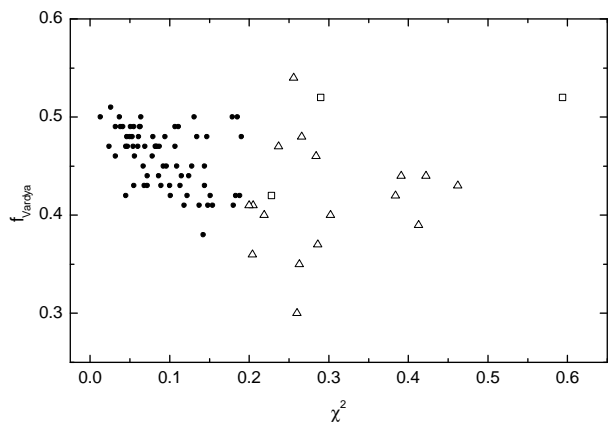}
\caption{Asymmetry parameter $f$ plotted against our difference parameter $\chi^{2}$. Above $\chi^{2}$=0.2
stars are separated into group 1 (open triangles) and 2 (open boxes) by the symbol used.}
\label{fvschi}
\end{figure}

To allow for an at least approximate comparison to older studies we compare in Fig.\,\ref{fvschi} 
the asymmetry parameter $f$ taken from Vardya (1988) with the value of $\chi^{2}$ derived in this study. 
We have 89 stars in common with the Vardya study. A nice relation of a decreasing value of $f$ with
an increasing value of $\chi^{2}$ can be clearly seen below $\chi^{2}$=0.2. 
Above that value the relation is somehow lost in a large scatter and in the low
number of available objects, but a fraction of the data points seem to still follow the relation
suggested from the objects with $\chi^{2}<$0.2. 
Stars are separated here into group 1 (open triangles) and group 2
(open boxes). We note that there is no
star in the sample of Vardya (1988) with a $f$-value below 0.3. On the other hand, for the most extreme
objects in our sample like DD Ara it is very difficult to determine the exact time of the minimum due to
its broadness.
We see also a branch of stars around $f$=0.5. These are stars, which 
are symmetric but not sinusoidal. Also the star with the largest value of $\chi^{2}$ in this comparison
belongs to this group. 
These findings reveal an advantage of our method in comparison with the older approach, namely the ability to
identify and characterize not only asymmetric but any non-sinusoidal light variation.

\section[]{Discussion}
Approximately 30\,\% of the mira light curves deviate significantly from a sinusoidal variation (see Fig.\,\ref{chi2distrib}). This value is rather independent from the inclusion of lower amplitude stars. If we
exclude all stars from our sample with a visual amplitude of less then 2.5\,mag, we find 28\,\% stars to
show a non-sinusoidal light change (compared to 33\,\% in the other case). This result is very similar to
the fraction of stars showing an asymmetry value $f$ outside the range 0.45$<f<$0.50 according to Vardya (1988). However, by using the whole light curve instead of only two points (maximum and minimum) we
think that our method leads to a more robust tool for studying the light curve shape circumventing the problem of
determining the minimum point in a rather flat and broad minimum. Furthermore, the 
group of stars
with $\chi^{2}$\,$>$\,0.2 includes also several stars with symmetric but non-sinusoidal light changes (group 2).
In combination with the two values $\chi_{1}$ and $\chi_{2}$,
defined as the absolute deviation of the normalized averaged light curve from the sinusoidal reference curve for the 
first and second half of the light change, 
two interesting classes of non-sinusoidal light curves could be detected
and well separated.

In the following we want to briefly investigate a possible connection between non-sinusoidal
light change and other stellar parameters. Two parameters directly result from the light curves, namely the stellar
period and the visual amplitude. None of the two parameters shows any clear correlation with the calculated values 
$\chi^{2}$, $\chi_{1}$ or $\chi_{2}$. We note, that Mennessier et al. (1997) come to the same result by using the
$f$ value (see their Fig.\,2).

\begin{figure}
\includegraphics[width=80mm]{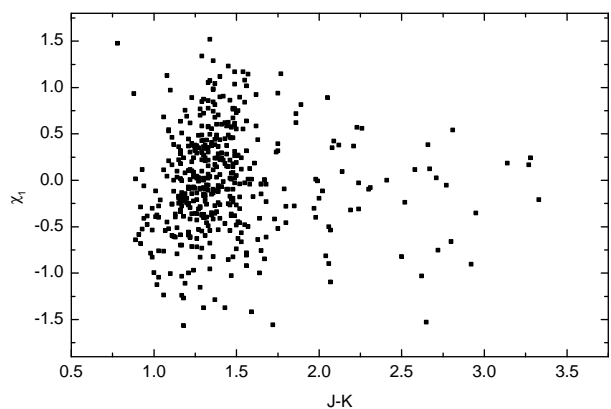}
\caption{Parameter $\chi_{1}$ vs.~$(J-K)$ from 2MASS.}
\label{chiJK}
\end{figure}

The ASAS catalogue further gives various colours for each object from
the 2MASS and the IRAS database. The colour of a sample star will depend both on the surface temperature and the
circumstellar reddening (plus some contribution from interstellar reddening, which was not taken into account here).
Both aspects are likely related to the mass loss, and from previous
studies described in the introductory section we would suspect to see a relation here. However, only a weak trend
of $\chi_{1}$ with $(J-K)$ could be found (Fig.\,\ref{chiJK}): stars with high (positive) values of 
$\chi_{1}$ (group 1) are not found among the reddest objects in our sample. Other colours tested, namely
$(V-K)$, $(V-$[12]$)$ and $($[12]$-$[25]$)$, all show no relation between any of our three parameters and 
colour. We note
that the variability of the objects will naturally introduce some scatter in the colour, which might affect 
this conclusion.

236 stars of our sample have a spectral type in the GCVS (linked via the variable name given in the ASAS catalogue). 
Of these, 23 (9\,\%) are C-stars (including 1 star classified as CS), 9 (3\,\%) are S-stars, two objects are
K-type stars, and the remaning stars are all of spectral type M. 
Due to the small fraction of C- and S-stars it is 
difficult to reach reliable conclusions on the relation between light curve shape and the atmospheric chemistry. 
Discriminating between sinusoidal and non-sinusoidal light variations based on a $\chi^{2}$ limit of 0.2, we find
that about 34\,\% of the S and C stars are showing non-sinusoidal variations, while only 25\,\% of the
oxygen-rich stars are found in this group. All S-stars with non-sinusoidal variations belong to group 2,
i.e. they show narrow minima and broad or double-peaked maxima. For the C-stars there is no clear trend
visible. 

\section[]{Conclusions}
Revisiting the question of the shape of mira light curves we find that about 30\,\% differ significantly
from a sinusoidal light change. A deviation in both directions is observed resulting in two main groups
of light curve shapes. While the study of Vardya (1988) focused only on asymmetric light curves (our group 1)
we present here for the first time statistics on the second group showing broad or even double peak maxima
(group 2). This kind of light change is observed almost as frequent as the well known asymmetric light curve.

We found no or only very weak
correlations with colour or light curve parameters like period or visual amplitude. 
The strongest correlation seems to exist
with atmospheric chemistry as S-type stars seem to deviate only into one direction (group 2). 
In general, S and C-stars show a higher fraction of non-sinusoidal variation
than the M-type stars. Based on our sample from the ASAS database we conclude that there is no
simple relation between the light curve shape and various observables tested here.  

Ongoing and future multiple epoch all-sky surveys will provide a wide potential area for applying the 
method described in this paper. These datasets will allow to re-investigate the question on which stellar
properties finally determine the shape of the light curve.
Interest on the typical light curve shape of miras comes from various sides, e.g.
to constrain dynamical model atmospheres for AGB stars (Nowotny, H\"ofner \& Aringer 2010), or to simulate mira
lightcurves in the preparation of data analysis software. This study was done in preparation of the 
variability analysis of data from the Gaia satellite.

\section*{Acknowledgments}
This research has been funded by the Austrian Science Fund (FWF) under project P20046-N16.
The author thanks Grzegorz Pojmanski and the anonymous referee for helpful comments.

\end{document}